\documentstyle[11pt,psfig]{article}
\def\be{\begin{equation}}
\def\ee{\end{equation}}
\def\bea{\begin{eqnarray}}
\def\eea{\end{eqnarray}}
\def\simlt{\stackrel{<}{{}_\sim}}
\def\simgt{\stackrel{>}{{}_\sim}}

\begin{document}
\begin{titlepage} 
\title{\bf{WHAT IS THE UPPER LIMIT ON THE LIGHTEST SUPERSYMMETRIC
HIGGS MASS?}\thanks{Work presented at the Sixth International Symposium
on Particles,
Strings and Cosmology (PASCOS-98), Northeastern University, Boston
MA 02115, USA, March 22-29 1998.}}
\vspace{1cm}
\author{MARIANO QUIROS\\
IEM (CSIC), Serrano 123, 28006-Madrid, Spain\\
E-mail:  quiros@pinar1.csic.es\\
\\
JOSE RAMON ESPINOSA\\
CERN TH-Division, CH-1211 Geneva 23,
Switzerland\\
E-mail: jose.espinosa@cern.ch}
\maketitle
\vspace{1.cm}   
\def\baselinestretch{1.15}
\begin{abstract}
In this talk the question of {\it what is
the upper bound on the lightest supersymmetric
Higgs mass, $m_h$} is addressed. This question is relevant 
since experimental lower bounds on $m_h$ might implement, 
in the near future, exclusion of supersymmetry.
By imposing (perturbative) unification of the gauge couplings
at some high scale $\simgt 10^{17}$ GeV, 
we have found that for a top-quark mass $M_t=175$ GeV, 
and depending on the supersymmetric parameters, 
this bound can be as high as 205 GeV.  
\end{abstract}  
\vspace{4cm}
\leftline{CERN-TH/98-292}
\leftline{September 1998}
 
\thispagestyle{empty}

\vskip-20.cm
\rightline{{ CERN-TH/98-292}}
\rightline{{ IEM--FT--180/98}}
\rightline{{ hep-ph/9809269}}
\vskip3in
 
\end{titlepage}

\section{Introduction: MSSM and gauge unification}

Low-energy supersymmetry (MSSM)~\cite{susy} is a key ingredient in the 
best-qualified candidate models to supersede the Standard Model (SM) at
energies beyond the TeV range. The MSSM is supported both by
theoretical arguments and experimental indirect hints. From the
theoretical point of view supersymmetry achieves cancellation of ultraviolet
quadratic divergences which is welcome to alleviate
the hierarchy problem. From the experimental side, using LEP
electroweak precision data, the MSSM achieves gauge coupling
unification at a value and scale given by 
\begin{equation}
\alpha_{\rm GUT}\sim 1/25,\quad 
M_{\rm GUT}\sim 2\times 10^{16}\,{\rm GeV}\ .
\end{equation}
On the other hand, the extensive experimental search of the
(super)-partners of SM elementary particles predicted by supersymmetry
(SUSY) has been unsuccessful so far, challenging~\cite{ft},
with the rise of experimental mass limits, the naturalness and relevance
of SUSY for electroweak-scale physics. 

In this context, the sector of the theory responsible for electroweak
symmetry breaking has a special status. While all superpartners of the
known Standard Model particles can be made heavy by simply 
rising soft supersymmetry-breaking
mass parameters in the model, the Higgs sector necessarily contains a
physical Higgs scalar whose mass does not depend sensitively on the
details of soft masses but is fixed by the scale of the electroweak symmetry
breaking\footnote{Here and throughout we assume symmetry breaking is
achieved in a weakly interacting Higgs sector. This is specially well
motivated and natural in SUSY models. The reader should be aware of other
more complicated possibilities \cite{Giudice}.}. 
More in detail, in the MSSM,
unlike in the SM, the quartic coupling is not an independent parameter but
\begin{equation}
\lambda=\frac{1}{2}(g_2^2+\frac{3}{5}g_1^2)\cos^2 2\beta \, ,
\label{lambdaMSSM}
\end{equation}
where $\tan\beta=\langle H_2^0\rangle/\langle H_1^0\rangle$ and
$g_2$ and $\sqrt{3/5}g_1$ are the $SU(2)_L\times U(1)_Y$ gauge couplings.

Equation (\ref{lambdaMSSM}) has the important consequence that
the value, and renormalization, of the quartic coupling in the
MSSM is related to that of $g_2^2+3g_1^2/5$. This means in turn
that the mass of one (usually the lightest one) of the CP-even
scalar mass eigenvalues is not controlled by the SUSY breaking
parameters, but by the electroweak breaking ($v=174.1$ GeV). The
existence of the SM-like Higgs is a generic feature of all
Supersymmetric Standard Models (SSM). In the MSSM
the bound
\be
m_h^2\leq M_Z^2\cos^22\beta\leq M_Z^2\, ,
\label{cota}
\ee
follows directly from (\ref{lambdaMSSM}). Eq.~(\ref{cota}) seems an
indication that non-discovery of the Higgs boson at LEP could
rule out the MSSM independently of any direct
SUSY searches. In fact, since present experimental bounds at LEP
already set fairly stringent bounds on the Higgs mass
\be
m_h\simgt 90\ {\rm GeV}\, ,
\ee
does it mean that the MSSM is already ruled out?

\section{Bounds in the MSSM}
Were it not for radiative corrections~\cite{radcor1,radcor2,radcor3} 
the MSSM would have already been excluded!
The leading effects are included in radiative corrections from
the top-stop sector, with $\widetilde{t}_{L,R}$ mixing parameterized by
\be
\widetilde{A}_t\equiv\frac{A_t-\mu/\tan\beta}{M_{\rm SUSY}}\, .
\label{mixing}
\ee

In the absence of mixing, leading higher-order 
corrections~\cite{rgmh1,rgmh2,rgmh3} 
are resummed by shifting
the energy scale ($M_{\rm SUSY}\rightarrow \sqrt{M_{\rm
SUSY}m_t}$) in the one-loop result\footnote{The physical top-quark mass
$M_t$ is related to the 
running $\overline{\rm MS}$ top-quark mass $m_t$ by the relation 
$m_t=M_t/[1+4\alpha_3(M_t)/3\pi]$, where $g_3$ is the $SU(3)$
coupling constant.}, 
\be
\Delta m^2_{h,rad}=\frac{3}{4\pi^2 v^2}m_t^4\left(\sqrt{M_{\rm
SUSY}m_t}\right) 
\log\frac{M_{\rm SUSY}^2}{m_t^2}\, .
\label{radcorr}
\ee
In the presence of mixing there is an important (threshold)
contribution as
\be
\Delta m^2_{h,th}=\frac{3}{8\pi^2}\frac{m_t^4}{v^2}
\widetilde{A}_t^2\left(2-\frac{1}{6}\widetilde{A}_t^2\right)\, .
\label{threshold}
\ee

These corrections push the upper mass limit for the lightest
Higgs boson of the MSSM up to $125$ GeV (for a top-quark 
mass $M_t=175$ GeV and $M_{SUSY}<1$ TeV), as can be seen from
the lowest curve in the right panel of Fig.~1.
So there appears an immediate question: what would happen if (when)
experimental bounds on the Higgs mass reach (and overcome) $\sim$ 125 GeV?
Should we give up SSMs?

\section{Bounds in the MSSM with a gauge singlet}

A singlet field $S$ added to the MSSM does not spoil unification and from
this point of view it is as legitimate as the MSSM itself. But
the $S$ field can couple in the superpotential $W$ to the
Higgs sector as 
\be
W=\lambda_1 S H_1\cdot H_2\, ,
\ee
leading to a contribution to the Higgs quartic coupling as
\be
\Delta\lambda=\lambda_1^2 \sin^2 2\beta\, .
\label{deltaS}
\ee
The contribution (\ref{deltaS}) can be sizable for small
values
of $\tan\beta$. An upper triviality  limit on $\lambda_1$ (and thus on
$\lambda(M_Z)$) exists if one requires perturbativity of  all parameters
of the theory in the energy range [$M_Z,M_{\rm GUT}$]. In fact the
absolute limit is reached when $\lambda_1(Q)$ goes
non-perturbative, 
i.e. when \footnote{The actual value on the
right hand side of Eq.~(\ref{nopert}) does not matter for the
precise value of the bound since the transition to the
non-perturbative regime is very abrupt.}
\be
\frac{\lambda_1^2(M_{\rm GUT})}{4\pi}={\cal O}(1)\, .
\label{nopert}
\ee
This constraint is very strong, since $\lambda_1$ increases
rapidly with the scale, and the gain in $m_h$ with respect to
the MSSM value is modest~\cite{eq92,eq93,gordy}.

Inspection of the  renormalization group equations~\cite{eq93}
\begin{eqnarray}
8\pi^2 \frac{d\lambda_1}{dt} &=&
\left[-\frac{3}{2}g_2^2-\frac{3}{10}g_1^2 +\frac{3}{2}(h_t^2+h_b^2)+
2\lambda_1^2 \right]\lambda_1 \nonumber\\
8\pi^2\frac{dh_{t}}{dt} &=& \left[-\frac{8}{3}g_3^2
-\frac{3}{2}g_2^2-\frac{13}{30}g_1^2+3h_t^2+\frac{1}{2}h_b^2+
\frac{3}{4} \lambda_1^2\right] h_t \nonumber\\
8\pi^2\frac{dh_{b}}{dt} &=& \left[-\frac{8}{3}g_3^2
-\frac{3}{2}g_2^2-\frac{7}{30}g_1^2+3h_b^2+\frac{1}{2}h_t^2+
\frac{3}{4} \lambda_1^2\right] h_b
\end{eqnarray}
shows that the presence of gauge couplings slows down the
evolution of the couplings $\lambda_1$, $h_{t,b}$ with the
renormalization scale. Therefore, the stronger the gauge
couplings, the slower the evolution of $\lambda_1$, the larger
values of $\lambda_1(M_Z)$ are consistent with perturbativity,
and finally the higher values of $m_h$ can be reached. Gauge
couplings can be strengthened, without spoiling gauge unification,
by adding complete (anomaly free) $SU(5)$ representations~\cite{alex}. 
The simplest possibility is adding a number $N_5$
of pairs $({\bf 5}+\bar{\bf 5})$. In fact for $N_5=4$
perturbativity is almost saturated ($\alpha_{\rm GUT}\sim 1/5$)
if $M_{5\bar 5}\sim M_{\rm SUSY}$ \footnote{Of course other
possibilities are accessible by introducing intermediate scales,
but the previous one will lead to the highest upper bounds.}.

The unification of the gauge couplings and the upper bounds on
$m_h$ for the MSSM plus a singlet and $N_5=4$ are shown in the
left and right panels, respectively, of Fig.~1.
\begin{figure}[htb]
\centerline{
\psfig{figure=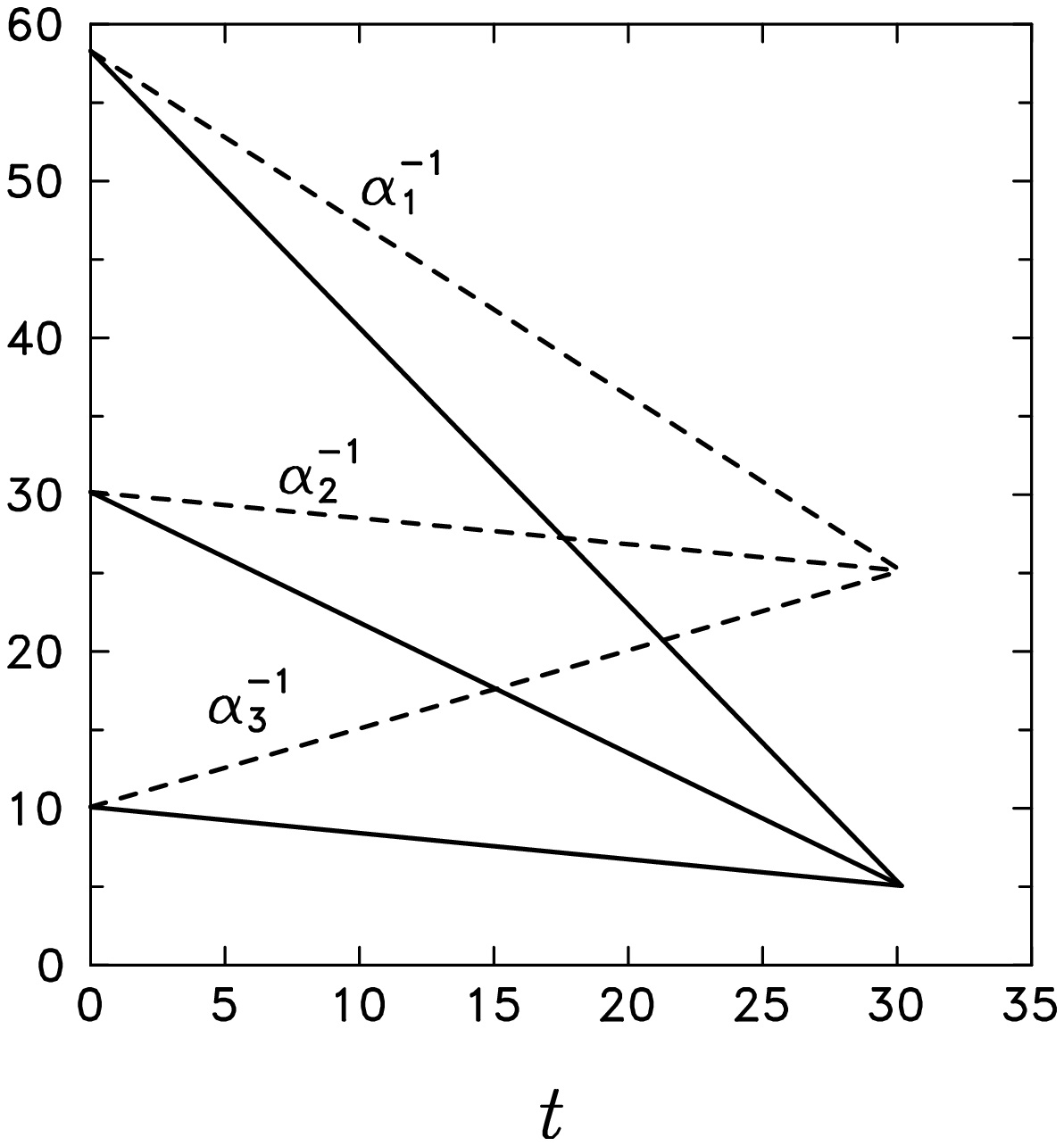,height=5.5cm,width=8cm,bbllx=-1.cm,bblly=3.cm,bburx=18.cm,bbury=17.cm}
\psfig{figure=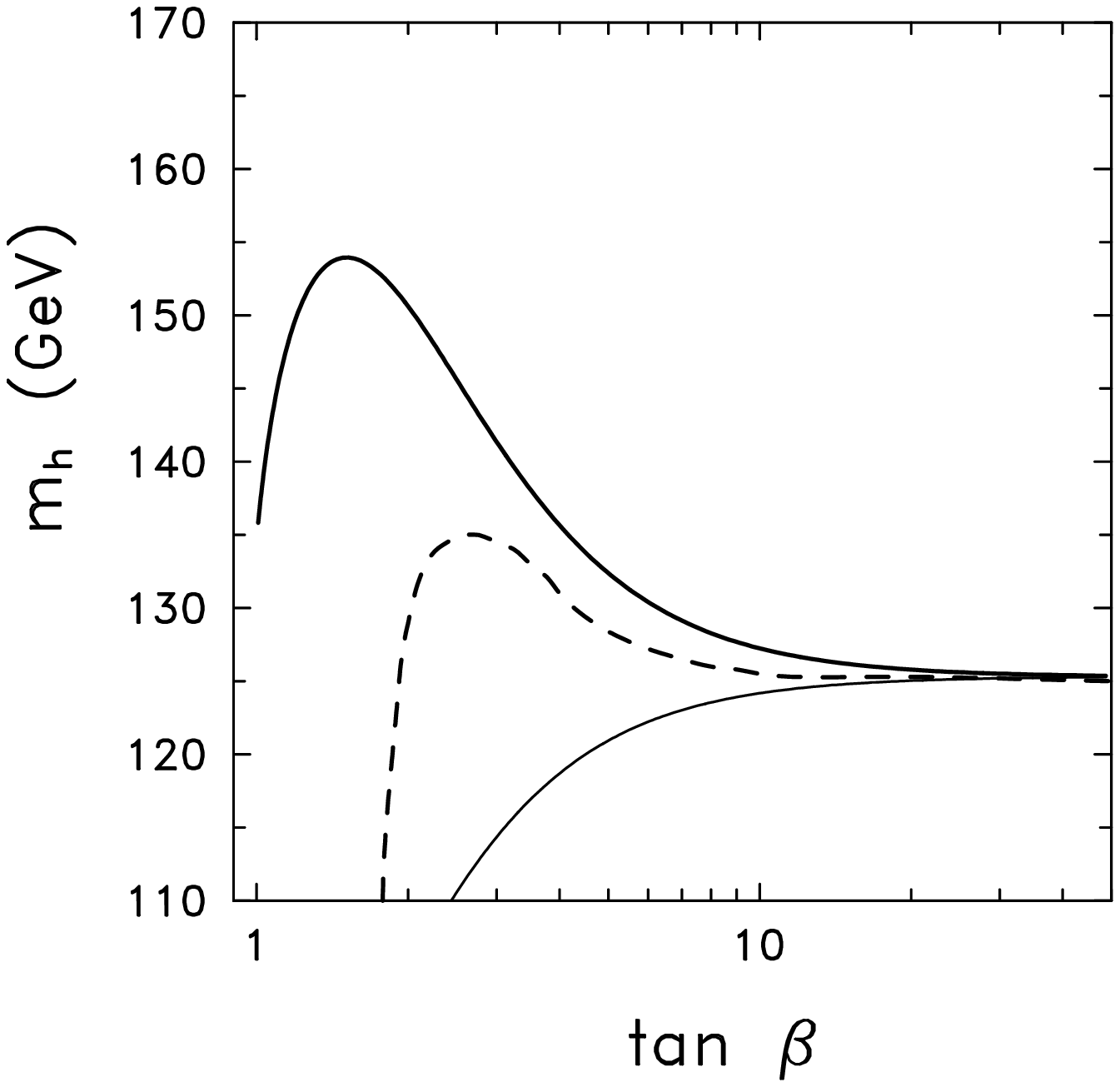,height=5.5cm,width=8cm,bbllx=2.5cm,bblly=3.cm,bburx=21.5cm,bbury=17.cm}}
\caption{Left panel: Running gauge couplings for the MSSM with
a gauge singlet (dashed lines) and the MSSM+4$({\bf 5}+{\bf \bar{5}})$ 
(solid lines) with $t=\log(Q/M_{SUSY})$. Right panel: upper limit on $m_h$ 
as a function of $\tan\beta$ for the MSSM (thin solid curve), 
the MSSM with a gauge singlet (thick dashed curve)
and the MSSM with a gauge singlet and $4({\bf 5}+{\bf \bar{5}})$
(thick solid line).}
\end{figure}
We observe from Fig.~1 that the maximum value of $m_h$ is at
$\tan\beta\sim 1.8$ and corresponds to $m_h\sim 155$ GeV, which
is a more sizable increase with respect to the MSSM.

Suppose again that our experimental colleagues exclude, in a few
years from now (LHC) a Higgs as light as $\sim$ 155 GeV! Should
we (once for all) exclude the SSM? We will next describe what we
consider to be the absolute upper bound on the lightest Higgs
mass in SSM consistent with gauge unification, or even with
perturbativity for scales below the Planck mass.

\section{Absolute bound and gauge unification}

To maximize the upper bound on $m_h$ we next assume that the model also
contains extra chiral multiplets with the appropriate quantum numbers to
give couplings of the form $W=h_X XH_iH_j$. Thus, $X$ can only be a
singlet ($S$) or a $Y=0,\pm1$ triplet ($T_Y$)~\cite{eq93,eq98}. Such terms
modify the quartic Higgs coupling via $F$-terms.
From the gauge-invariant trilinear superpotential 
\be
\label{superp}
W= \lambda_{1} H_1\cdot H_2 S
+  \lambda_{2} H_1\cdot T_0 H_2+  \chi_{1} H_1\cdot T_1 H_1   +
 \chi_{2} H_2\cdot T_{-1} H_2\, , 
\ee
the tree-level mass bound follows:
\bea
\label{bound}
m_h^2/v^2&\leq&\frac{1}{2}(g_2^2+\frac{3}{5}g_1^2)\cos^2 2\beta +
\left(\lambda_{1}^2
+\frac{1}{2}\lambda_{2}^2\right)\sin^2 2\beta\nonumber\\
&+& 4\chi_1^2\cos^4\beta+ 4\chi_2^2\sin^4\beta\, .
\eea
Different terms have different $\tan\beta$ dependence  and are
important in different regimes. For example, the $\chi_2$
contribution will be crucial for the upper limit
in the large-$\tan\beta$ region.
$S$ and $T_0$ induce the
same $\tan\beta$ dependence but the $\lambda_1$ correction can 
be more important than that of $\lambda_2$. For this reason we do
not take into account the possible effect of $T_0$ representations.

To achieve unification with only one scale $M_{SUSY}$ ($=$ 1 TeV) is
not completely trivial. When the MSSM is enlarged by one singlet $S$
and a pair $\{T_1,T_{-1}\}$ (to cancel anomalies) the running $g_1^2$ 
and $g_2^2$ meet at $M_X\sim 10^{17}$ GeV. Interestingly
enough, this is closer to the heterotic string scale than the 
MSSM unification scale. Of course,
$g_3^2$ fails to unify unless extra matter is added. This can be
achieved, for example, by adding 4 ($3+\bar{3}$) [$SU(2)_L\times U(1)_Y$
`singlet quark' chiral multiplets] or ($3+\bar{3}$) plus one
$SU(3)_c$ octet. In addition to this, we can still have a 
($5+\bar{5}$) $SU(5)$ pair, which will not change the unification scale. 
The unification of the couplings is shown in Fig.~2 (left panel,
solid lines).

For comparison, dashed lines show the
running couplings when their RG beta functions are fixed in such a way
that
all couplings reach a Landau pole at the unification scale. 
In this case the low-energy couplings are fully 
determined by the `light' matter content of
the model (which fixes the RG beta functions).
The
dashed lines can be considered as the perturbative upper limit on the
gauge couplings, and comparison with the solid lines shows that our model
is close to saturation and represents a concrete realization of the most
extreme scenario to maximize the $m_h$ bound. The emphasis here should lie
in this fact rather than in the plausibility or physics motivation of the
model {\it per se}. 

Having optimized in this way the most appropriate running gauge
couplings, we
turn to the running of $\lambda_1$ and $\chi_{1,2}$. The relevant RG
equations are~\cite{eq92,eq93}:
\begin{eqnarray}
8\pi^2 \frac{d\lambda_1}{dt} &=&
\left[-\frac{3}{2}g_2^2-\frac{3}{10}g_1^2 +\frac{3}{2}(h_t^2+h_b^2)+
3(\chi_1^2+\chi_2^2)+2\lambda_1^2 \right]\lambda_1 \nonumber\\
8\pi^2 \frac{d\chi_{1,2}}{dt} &=&
\left[-\frac{7}{2}g_2^2-\frac{9}{10}g_1^2+3h_{b,t}^2
+7\chi_{1,2}^2+\lambda_1^2 \right]\chi_{1,2} \nonumber\\
8\pi^2\frac{dh_{t}}{dt} &=& \left[-\frac{8}{3}g_3^2
-\frac{3}{2}g_2^2-\frac{13}{30}g_1^2+3h_t^2+\frac{1}{2}h_b^2+
3\chi_2^2+
\frac{1}{2} \lambda_1^2\right] h_t \nonumber\\
8\pi^2\frac{dh_{b}}{dt} &=& \left[-\frac{8}{3}g_3^2
-\frac{3}{2}g_2^2-\frac{7}{30}g_1^2+3h_b^2+\frac{1}{2}h_t^2+
3\chi_2^2+ \frac{1}{2} \lambda_1^2\right] h_b\, .
\end{eqnarray}
For a given value of $\tan\beta$ (which
influences the top and bottom Yukawa couplings entering the RGs)
we find the maximum value of the particular combination of Yukawa
couplings that enters the bound (\ref{bound}), compatible with
perturbativity up to $M_S$.
To add the important radiative corrections we follow
Refs.~\cite{rgmh1,rgmh2,rgmh3},
which include two-loop RG improvement and stop-mixing effects.
We fix  $M_t=175$ GeV.

\begin{figure}[htb]
\centerline{
\psfig{figure=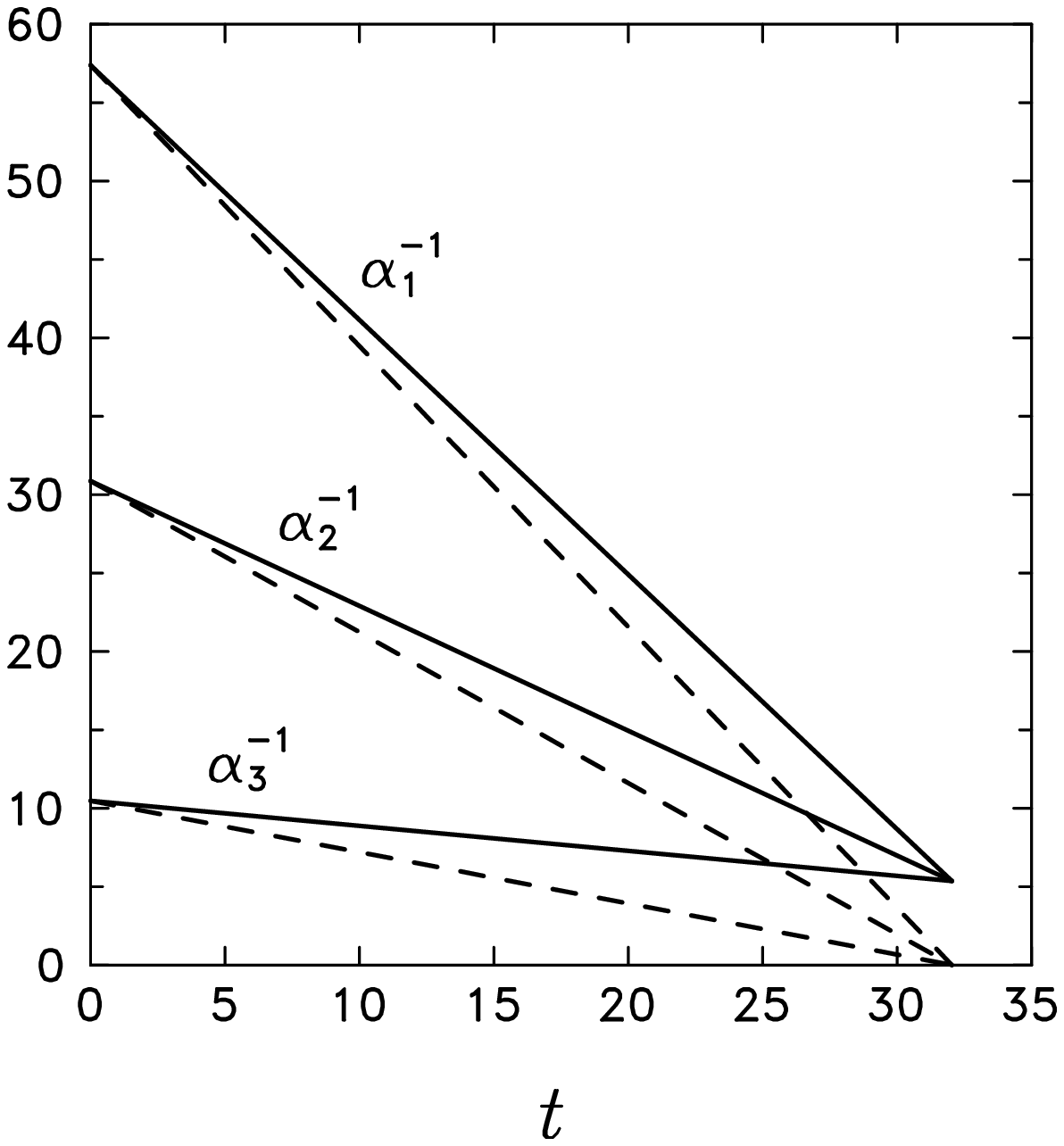,height=5.5cm,width=8cm,bbllx=-1.cm,bblly=3.cm,bburx=18.cm,bbury=17.cm}
\psfig{figure=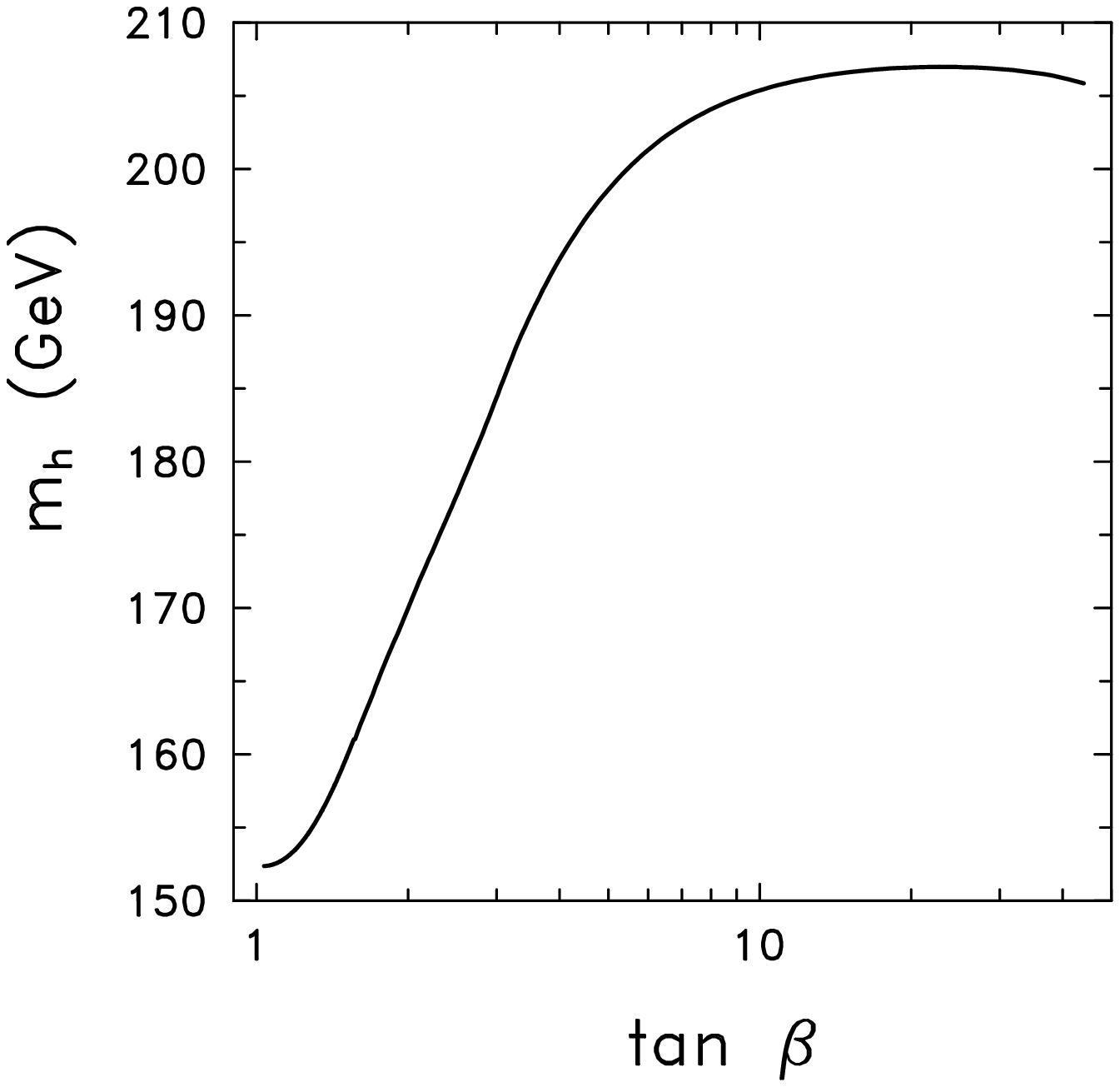,height=5.5cm,width=8cm,bbllx=2.5cm,bblly=3.cm,bburx=21.5cm,bbury=17.cm}}
\caption{Left panel: running gauge couplings for the model
discussed in the text (solid lines) and its upper
perturbative limit (dashed lines). Right panel: Upper bound on
$m_h$ for both models.}
\end{figure}

Before presenting our results, it is worth discussing in more
detail the bound presented in Eq.~(\ref{bound}). If the extra fields
responsible
for the enhancement of $m_h$ sit at 1~TeV, should their effect not
decouple from the low-energy effective theory? Indeed, 
in a simple toy model with
\be
W=mS^2+\lambda_1 H_1\cdot H_2 S\, ,
\ee
the limit $m\gg M_Z, M_{\rm SUSY}$ leads to an effective superpotential
in which no renormalizable coupling of $S$ to 
$H_1\cdot H_2$ exists [technically, the $F$-term contribution 
($\sim \lambda_1^2$) to the Higgs doublet
self-interactions is cancelled by a tree diagram that interchanges the
heavy singlet, thus realizing decoupling]. If, to avoid this, we
take $m\sim M_Z$, more than one light Higgs will appear in the spectrum.
A complicated mixed squared-mass matrix results whose lightest
eigenvalue
does not saturate the bound (\ref{bound}). Is then this mass limit simply
too conservative an overestimate of the real upper limit?
It is easy to convince oneself that, in the presence of soft breaking
masses, the perfect decoupling cancellation obtained in the large SUSY
mass limit does not take place (we assume $m\simlt M_{\rm SUSY}$) 
and the final lightest Higgs mass depends in a complicated way on these
soft mass parameters. The interesting outcome is that soft masses can be
adjusted in order to saturate the bound (\ref{bound}); the numbers
we will present can thus be reached in particular models and no limits
lower than these can be given without additional assumptions (which 
we will not make here, in the interest of generality).

The final bound, with radiative corrections included, is presented in
Fig.~2 (right panel). For $\tan\beta \simlt 2$ the main
contribution to the $m_h$ value comes from the $\lambda_1$
coupling. In particular for $\tan\beta\simeq 1.8$ the effect of
$\lambda_1$ is maximized and yields a bound on $m_h\sim$ 155
GeV, as in the model presented in the previous section~\cite{alex}. 
For larger values of $\tan\beta$ the effect of
$\lambda_1$ drops off and $\chi_2$ dominates. We also see that the bound
can reach values as large as
205-210 GeV, that we should consider as the absolute bound in
supersymmetric models, which are perturbative up to the high
scale and without intermediate scales. Finally we can notice
that the $\chi_1$ coupling does play a role only for 
$\tan\beta \sim 1$~\cite{eq98} where the bound is $\sim$ 140 GeV.

\section{Conclusions} 

In conclusion, we calculate a numerical absolute upper limit on
the mass of the lightest supersymmetric Higgs boson for any model 
with arbitrary matter content compatible with gauge coupling unification 
around (and perturbativity up to) the
string scale. With this assumption, we show that this light Higgs mass 
can be as high as $\sim$ 205 GeV. 
The model saturating this bound has asymptotically divergent
gauge couplings and points toward non-perturbative unification. 
Besides being of obvious interest to the experimentalists, this
result has interest for theorists too. If Higgs searches reach the MSSM
bounds without finding a signal for a Higgs boson, this could be taken,
if one is willing to stick to (perturbative) low-energy supersymmetry, as
evidence for additional matter beyond the minimal model.

\end{document}